\documentclass{article}
\usepackage{amssymb}
\usepackage{amsmath}

\begin{document}

\bigskip

\begin{center}
{\Large Some connections between $BCK$-algebras and }$n-${\Large ary block
codes}

\begin{equation*}
\end{equation*}

A. Borumand Saeid, Cristina Flaut, \ Sarka Ho\v{s}kov\'{a}-Mayerov\'{a},\
Roxana-Lavinia Cristea, M. Afshar, M. Kuchaki Rafsanjani%
\begin{equation*}
\end{equation*}
\end{center}

\textbf{Abstract.} {\small In the last time some papers were devoted to the
study of the connections between binary block codes and BCK-algebras.} 
{\small In this paper, we try to generalize these results to $n$-ary block
codes, providing an algorithm which allows us to construct a $BCK$-algebra
from a given $n$-ary block code.\bigskip }

\bigskip


\begin{equation*}
\end{equation*}

\textbf{0. Introduction}

\begin{equation*}
\end{equation*}

\ Y. Imai and K. Iseki introduced $BCK$-algebras \ in 1966, through the
paper [Im, Is; 66], as a generalization \ of the concept of set-theoretic
difference and propositional calculi. This class of \ $BCK$-algebras is a
proper subclass of the class of $BCI$-algebras and has many applications to
various domains of mathematics.

One of the recent applications of $BCK$-algebras was given in the Coding
Theory. In the paper [Ju,So; 11], the authors constructed a finite binary
block-codes associated to a finite $BCK$-algebra. In [Fl; 15], the author
proved that, in some circumstances, the converse of the above statement is
also true and in the paper [B,F; 15] the authors proved that binary block
codes are an important tool in providing orders with which we can build
algebras with some asked properties. For other details regarding $BCK$%
-algebras, the reader is referred to [Is, Ta; 78].

In general, the alphabet on which are defined block codes are not binary. It
is used an alphabet with $n$ elements, $n\geq 2,$ identified usually with
the set $A_{n}=\{0,1,2,...,n-1\}.$ These codes are called $n-$ary block
codes. In the present paper, we will generalize this construction of binary
block codes to $n-$ary block codes. For this purpose, we will prove that to
each $n-$ary block code $V$ we can associate a $BCK$-algebra $X$ such that
the $n-$ary block-code generated by $X,V_{X},$ contains the code $V$ as a
subset.%
\begin{equation*}
\end{equation*}

\begin{equation*}
\end{equation*}

\begin{center}
\bigskip
\end{center}

\textbf{1. Preliminaries}

\begin{equation*}
\end{equation*}

\textbf{Definition 1.1.} An algebra $(X,\ast ,\theta )$ of type $(2,0)$ is
called a \textit{BCI-algebra} if the following conditions are fulfilled:

$1)~((x\ast y)\ast (x\ast z))\ast (z\ast y)=\theta ,$ for all $x,y,z\in X;$

$2)~(x\ast (x\ast y))\ast y=\theta ,$ for all $x,y\in X;$

$3)~x\ast x=\theta ,$ for all $x\in X$;

$4)$ For all $x,y,z\in X$ such that $x\ast y=\theta ,y\ast x=\theta ,$ it
results $x=y$.\newline

If a $BCI$-algebra $X$ satisfies the following identity:

$5)$ $\theta \ast x=\theta ,~$for all $x\in X,$ then $X$ is called a \textit{%
BCK-algebra}.\newline

A $BCK$-algebra $X$ is called \textit{commutative }if $x\ast (x\ast y)=y\ast
(y\ast x),$ for all $x,y\in X$ and \textit{implicative }if $x\ast (y\ast
x)=x,$ for all $x,y\in X.$ A $BCK$-algebra $(A,\ast ,0)$ is called \textit{%
positive implicative} if and only if%
\begin{equation*}
\left( x\ast y\right) \ast z=\left( x\ast z\right) \ast (y\ast z),\text{ for
all }x,y,z\in A.
\end{equation*}

The partial order relation "$\leq $" on a $BCK$-algebra is defined such that 
$x\leq y$ if and only if $x\ast y=\theta .$

An equivalent definition of $BCK$-algebra was gave in the following
proposition.\medskip

\textbf{Proposition 1.2.} ([Me, Ju; 94], Theorem 1.6) \ \textit{An algebra} $%
(X,\ast ,\theta )$ \textit{of type} $(2,0)$ \textit{is a BCK-algebra if and
only if the following conditions are satisfied:}

1) $((x\ast y)\ast (x\ast z))\ast (z\ast y)=\theta ,$ \textit{for all} $%
x,y,z\in X;$

2) $x\ast \left( 0\ast y\right) =x,$\textit{for all} $x,y\in X;$

3) \textit{For all} $x,y,z\in X$ \textit{such that} $x\ast y=\theta ,y\ast
x=\theta ,$ \textit{it results} $x=y$.\medskip

Let $(X,\ast ,\theta )$ be a finite $BCK$-algebra with $n$ elements and $A$\
be a finite nonempty set. A map $f:A\rightarrow X$ is called \textit{a
BCK-function}. Let $A_{n}=\{0,1,2,...,n-1\}$. In the following, we will
consider $BCK$ algebra $X$ and the set $A$ under the form: $%
X=\{r_{0},r_{1},...,r_{n-1}\},A=\{x_{0},x_{1},...,x_{m-1}\},m\leq n.$ A 
\textit{cut function} of $f$ \ is a map $f_{r_{j}}:A\rightarrow A_{n},$ $%
r_{j}\in X,$ such that $f_{r_{j}}\left( x_{i}\right) =k$ if and only if $%
r_{j}\ast f\left( x_{i}\right) =r_{k},$ for all $r_{j},r_{k}\in X,x_{i}\in
A,i,j,k\in \{0,1,2,...,n-1\}.$ For each $BCK$-function $f:A_{n}\rightarrow X$%
, we can define an $n-$ary block-code with codewords of\ length $m$. For
this purpose, we consider to each element $r\in X$ the cut function $%
f_{r}:A\rightarrow A_{n},r\in X.$ To each such a function, will correspond
the codeword $w_{r},$ with symbols from the set $A_{n}.$ We have $%
w_{r}=w_{0}w_{1}...w_{n-1},$ with $w_{i}=j,$ $j\in A_{n},~$if and only if \ $%
f_{r}\left( x_{i}\right) =j,$ that means $r\ast f\left( i\right) =r_{j}.$ We
denote this code with $V_{X}.$ In this way, we can associate to each $BCK$%
-algebra an $n-$ary block code.\medskip

\textbf{Example 1.3.} We consider the following $BCK$-algebra $(X,\ast
,\theta ),$ with the multiplication given in the following table (see
[Ju,So; 11] , Example 4.2).

\begin{tabular}{l|llll}
$\ast $ & $\theta $ & $a$ & $b$ & $c$ \\ \hline
$\theta $ & $\theta $ & $\theta $ & $\theta $ & $\theta $ \\ 
$a$ & $a$ & $\theta $ & $\theta $ & $a$ \\ 
$b$ & $b$ & $a$ & $\theta $ & $b$ \\ 
$c$ & $c$ & $c$ & $c$ & $\theta $%
\end{tabular}%
\newline
We have $X=\{\theta ,a,b,c\}~A=A_{4}=\{0,1,2,3\}.$We consider $%
f:A\rightarrow X,f\left( 0\right) =\theta ,f\left( 1\right) =a,f\left(
2\right) =b,f\left( 3\right) =c$ and $f_{r}:A_{4}\rightarrow A_{4},r\in X,$
a cut function.

To $\ r=\theta ,$ corresponds the codeword $w_{\theta }=0000.$ For $r=a,$ we
obtain the codeword $1001$. Indeed, $f_{a}\left( 0\right) =1,$ since $a\ast
f\left( 0\right) =a\ast \theta =a=f\left( 1\right) ;f_{a}\left( a\right) =0$
since $a\ast f\left( 1\right) =a\ast a=\theta =f\left( 0\right) ;f_{a}\left(
b\right) =0$ and $a\ast f\left( 2\right) =a\ast b=\theta =f\left( 0\right)
;~f_{a}\left( c\right) =1$, also $a\ast f\left( 3\right) =a\ast c=a=f\left(
1\right) ;$\newline

We wonder if and in what circumstances the converse is also true?

In the following, we will try to find answers at this question.

\begin{equation*}
\end{equation*}

\textbf{2. Main results}\newline
\begin{equation*}
\end{equation*}

Let $A_{n}^{\prime }=\{1,2,...,n-1\}$ be a finite set and $%
V=\{w_{1},w_{2},...,w_{m}\}$ be $\ n-$ary codewords, ascending ordered after
lexicographic order. We consider $w_{i}=w_{i1}w_{i2}...w_{iq},w_{ij}\in
A_{n}^{\prime },$ $j\in \{1,2,...,q\},$ with $w_{ij}$ descending ordered
such that 
\begin{equation*}
w_{iw_{ik}}\leq k,~~i\in \{1,2,...,m\},~~k\in \{1,2,...,\min \{n-1,q\}\}
\end{equation*}%
and $w_{ij}=1$ in the rest.\newline

\textbf{Definition 2.1.} Let $V$ be the $n-$ary codeword, defined above. To
this code we associate a matrix $M=\left( \alpha _{st}\right) _{s,t\in
\{0,1,...,r-1\}},$ $M\in \mathcal{M}_{r}\left( A_{n}\right) ,$ where $r$ is
defined in the following.\medskip

\textbf{Case 1. }$q<n.$ Let $r=n-1+m.$ We define $\alpha _{ss}=0,$ $\alpha
_{s0}=s,\alpha _{0s}=0,$ $s\in \{0,1,2,...,r-1\}.$ For $1\leq s\leq n-1,$
put $\alpha _{st}=$ $1,$ if $t\leq s,$ $\alpha _{st}=0,$ if $t\geq s.$ For $%
\ s\geq n-1,$ define $\alpha _{st}=w_{it},$ for $t\in \{1,2,...,q\}$ and $%
\alpha _{sq+j}=1,$ for $q+j<s.$ We have $\alpha _{st}=0,$ for $t\geq s.$

\textbf{Case 2.} $q\geq n.$ Let $r=m+q+1.$ We define $\alpha _{ss}=0,$ $%
\alpha _{s0}=s,\alpha _{0s}=0,s\in \{0,1,2,...,r-1\}.$ For $1\leq s\leq q,$
define $\alpha _{st}=$ $1,$ if $t\leq s,$ $\alpha _{st}=0,$ if $t\geq s.$
For $\ s>q,$ put $\alpha _{st}=w_{it},$ for $t\in \{1,2,...,q\}$ and $\alpha
_{sq+j}=1,$ for $q+j<s$. We have $\alpha _{st}=0,$ for $t\geq s.\medskip $

The matrix $M$ is called \textit{the matrix associated to the }$n-$\textit{%
ary block code} $V=\{w_{1,}w_{2},...,w_{m}\}$ and is a lower triangular
matrix. Example of such a matrix can be found in Section 3.\medskip 

\textbf{Definition 2.2.} With the above notations, let \ $M\in \mathcal{M}%
_{r}\left( A_{n}\right) $ $\ $ be the matrix associated to the $n-$ary block
code $V=\{w_{1,}w_{2},...,w_{m}\}$ defined on $A_{n}^{\prime }~$and $%
A_{r}=\{0,1,...,r-1\}$ be a nonempty set. We define on $A_{r}$ the following
multiplication 
\begin{equation*}
i\ast j=\alpha _{ij}=w_{ij}=k.\medskip
\end{equation*}

\textbf{Theorem 2.3.} \textit{With the above notations, we have that }$%
\left( A_{r},\ast ,0\right) $ \textit{is a BCK-algebra.\medskip }

\textbf{Proof.} \ Since conditions 2), 3) from Proposition 1.2 \ are
satisfied using Definition 2.1, we will only prove that $(\left( i\ast
j\right) \ast \left( i\ast k\right) )\ast \left( k\ast j\right) =0,$ for all 
$i,j,k\in \{0,1,...,r-1\}.$

\textbf{Case 1: }$\mathbf{j=0,k\neq 0}.$ We will prove that $\left( i\ast
\left( i\ast k\right) \right) \ast k=0.$ For $i=0$ it is clear.

For $k=0,$ we obtain $\left( i\ast \left( i\ast 0\right) \right) \ast
0=(i\ast i)\ast 0=0.$

For $k\neq 0,$ $i\geq r-m,k\in \{1,2,...,q\},$we have $\left( i\ast \left(
i\ast k\right) \right) =w_{iw_{ik}}\leq k,$ therefore $\left( i\ast \left(
i\ast k\right) \right) \ast k=0.$

For $k\neq 0,$ $i\geq r-m,k\geq q+1,i\geq k,$we have \ $\left( i\ast \left(
i\ast k\right) \right) \ast k=0,$ since $i\ast k=1,i\ast 1\leq n-1<k.$

For $i<r-m,$ $k\leq q+1,$ we have $\left( i\ast \left( i\ast k\right)
\right) \ast k=0$ since $i\ast k=1,i\ast 1=1$ and $1\ast k=0.$

For \ $i<r-m,$ $k>q+1,$ we have $\left( i\ast \left( i\ast k\right) \right)
\ast k=0$ since $i\ast k=0,$ we obtain $\left( i\ast 0\right) \ast k=i\ast
k=0.$

\textbf{Case 2}$\mathbf{:k=0,j\neq 0.}$ We will prove that $\left( i\ast
j\right) \ast i=0.$ We always have that $i\ast j\leq i,$ therefore $\left(
i\ast j\right) \ast i=0.$

\textbf{Case 3:} $\mathbf{k\neq 0,j\neq 0.}$ We will prove that $(\left(
i\ast j\right) \ast \left( i\ast k\right) )\ast \left( k\ast j\right) =0.$
For $i=0,$ it is clear. We suppose that $i\neq 0.$

For $i\geq r-m$ and $j,k<r-m,j<k.$ We have $n-1\geq \left( i\ast j\right)
\geq \left( i\ast k\right) ,$ therefore $(\left( i\ast j\right) \ast \left(
i\ast k\right) )=1.$ We also obtain $k\ast j=1,$ therefore $(\left( i\ast
j\right) \ast \left( i\ast k\right) )\ast \left( k\ast j\right) =1\ast 1=0.$

For $i\geq r-m$ and $j,k<r-m,k<j.$ We have $n-1\geq \left( i\ast j\right)
\leq \left( i\ast k\right) ,$ therefore $(\left( i\ast j\right) \ast \left(
i\ast k\right) )=0.$ It results that $(\left( i\ast j\right) \ast \left(
i\ast k\right) )\ast \left( k\ast j\right) =0.$

For $i\geq r-m$ and $j,k\geq r-m,j<k.$ We can have $i\ast j=1$ and $i\ast
k=1,$therefore $\left( i\ast j\right) \ast \left( i\ast k\right) =0.\,$\ We
can also have $i\ast j=1,i\ast k=0$ and $k\ast j=1,$ since $j<k.$ It results
that $(\left( i\ast j\right) \ast \left( i\ast k\right) )\ast \left( k\ast
j\right) =(1\ast 0)\ast 1=1\ast 1=0.$ Or, we can have $i\ast j=0,i\ast k=0,$
therefore the asked relation is zero.

For $i\geq r-m$ and $j,k\geq r-m,k<j.$ We can have $i\ast j=1$ and $i\ast
k=1,$therefore $\left( i\ast j\right) \ast \left( i\ast k\right) =0.$ \ Or,
we can have $i\ast k=1,i\ast j=0$ and $k\ast j=0,$ therefore we obtain zero.
We also can have $i\ast j=0,i\ast k=0,$ therefore the asked relation is zero.

For $i\geq r-m$ and $k<r-m<j.$ We can have $i\ast j=0,$ therefore the asked
relation is zero. We can have $i\ast j=1.$ It results $(\left( i\ast
j\right) \ast \left( i\ast k\right) )\ast \left( k\ast j\right) =\left(
1\ast \left( i\ast k\right) \right) \ast 0=1\ast \beta =0,$ since $k<j$ and $%
\beta \geq 0.$

For $i\geq r-m$ and $j<r-m<k.$ We have $i\ast j=1.$If $i\ast k=1,$ we obtain
zero. $\ $If $i\ast k=0,$ it results $(\left( i\ast j\right) \ast \left(
i\ast k\right) )\ast \left( k\ast j\right) =\left( 1\ast 0\right) \ast
\left( k\ast j\right) =1\ast \left( k\ast j\right) =0,$ since $k\ast j\geq
1. $

For $i<r-m$ and $j,k<r-m,j<k.$ $\ $ We have $i\ast j=1,i\ast k=1,$ therefore
we obtain zero.

For $i<r-m$ and $j,k<r-m,k<j.$ We can have $(\left( i\ast j\right) \ast
\left( i\ast k\right) )\ast \left( k\ast j\right) =\left( 1\ast 1\right)
\ast 0=0.$ Or, we can have $\left( i\ast j\right) =0,$ therefore we obtain
zero.

For $i<r-m$ and $j,k<r-m,j<n-1+\max \{q,m\}-m\leq k.$ We have $i\ast
j=1,i\ast k=0$ and $k\ast j=1.$ It results $(\left( i\ast j\right) \ast
\left( i\ast k\right) )\ast \left( k\ast j\right) =\left( 1\ast 0\right)
\ast 1=1\ast 1=0.$

For $i<r-m$ and $k<r-m,k<r-m\leq j.$ We can have $(\left( i\ast j\right)
\ast \left( i\ast k\right) )\ast \left( k\ast j\right) =\left( 1\ast
1\right) \ast 0=0.$ Or, we can have $\left( i\ast j\right) =0,$ therefore we
obtain zero.

For $i<r-m$ and $j,k\geq r-m,j<k.$ We have $\left( i\ast j\right) =0,$
therefore we obtain zero.

For $i<r-m$ and $j,k\geq r-m,j>k.$ We have $\left( i\ast j\right) =0,$
therefore we obtain zero. $\Box \medskip $

\textbf{Remark 2.4. }

1) $BCK$-algebra $\ \left( A_{r},\ast ,0\right) $ obtained in Theorem 2.3 is
unique up to an isomorphism.

2) From Theorem 2.3, let $\ \left( A_{r},\ast ,0\right) $ be the obtained $%
BCK$-algebra, with $A_{r}=\{0,1,2,...r-1\}.$ If $X=\{a_{0}=\theta
,a_{1},a_{2},...,a_{r-1}\},$ with multiplication "$\circ $" given by the
relation $a_{i}\circ a_{j}=a_{k}$ if and only if $i\ast j=k,$ for $i,j.k\in
\{0,1,2,...,r-1\},$ then $\left( X,\circ ,\theta \right) $ is a $BCK$%
-algebra.

3) If we consider $A_{q}=\{0,1,2,...q-1\},$ the map $f:A_{q}\rightarrow
X,f\left( i\right) =a_{i},$ gives us a code $V_{X},$ associated to the above 
$BCK$-algebra $\left( X,\circ ,\theta \right),$ which contains the code $V$
as a subset.\medskip

\textbf{Definition 2.5.} Let $(X,\ast ,\theta )$ be a $BCK$-algebra, and $%
I\subseteq X.$ We say that $I$ is a \textit{right-ideal} for the algebra $X$
if $\theta \in I$ and $x\in I,y\in X$ imply $x\ast y\in I$. An ideal $I$ of
a $BCK$-algebra $X$ is called \textit{a closed ideal }if it is also \textit{a%
} \textit{subalgebra }of $X$ (i.e. $\theta \in I$ and if $x,y\in I$ it
results that $x\ast y\in I$).\medskip

Let $V$ be an $n-$ary block code. From Theorem 2.3 and Remark 2.4, we can
find a $BCK$-algebra $X$ such that the obtained $n-$ary block-code $V_{X}$
contains the $n-$ary block-code $V$ as a subset.

Let $V$ be a binary block code with $m$ codewords of length $q.$ With the
above notations, let $X$ \ be the associated $BCK$-algebra and \ $W$\ $%
=\{\theta ,w_{1},...,w_{r}\}~$the associated $n-$ary block code which
include the code $V.$ We consider the codewords $\theta
,w_{1,}w_{2},...,w_{r}$ lexicographically ordered, $\theta \geq
_{lex}w_{1}\geq _{lex}w_{2}\geq _{lex}...\geq _{lex}w_{r}.$ Let $M\in 
\mathcal{M}_{r}\left( A_{n}\right) $ be the associated matrix with the rows $%
\theta ,w_{1},...,w_{r},$ in this order. Let $L_{w_{i}}$ and $C_{w_{j}}$ be
the lines and columns in the matrix $M$. We consider the sub-matrix $%
M^{\prime }$ of the matrix $M$ with the rows $L_{w_{1}},...,L_{w_{m}}$ and
the columns $C_{w_{m+1}},...,C_{w_{m+q}},~$which is the matrix associated to
the code $C.$ \medskip

\textbf{Proposition 2.6.} \textit{With the above notations, we have that}%
\newline
\textit{\ }$\{\theta ,w_{1},w_{r-m},w_{r-m+1},...,w_{r}\}$\textit{\
determines a closed right ideal in the algebra} $X.\medskip $

\textbf{Proof.} Let $Y=$ $\{\theta ,w_{1},w_{r-m},w_{r-m+1},...,w_{r}\}.$ We
will prove that $y\in Y,x\in X$ imply $y\ast x\in Y.$ From the definition of
the multiplication in the algebra $X,$ we have that $y\ast x\in \{\theta
,w_{1}\}.$ In the same time, if $x,y\in Y,$ it results that $x\ast y\in Y,$
since $y\ast x\in \{\theta ,w_{1}\}.$ 

\bigskip 
\begin{equation*}
\end{equation*}

\bigskip \textbf{3.} \textbf{Examples}%
\begin{equation*}
\end{equation*}

\textbf{Example 3.1.} Consider $A_{7}=\{0,1,2,3,4,5,6\},$ $n=7,$ $q=4,$ $%
m=3, $ $r=9,$ $V=\{w_{1},w_{2},w_{3}\},$ with $%
w_{1}=3211,w_{2}=4221,w_{3}=4321.$ The matrix $M$ associated to the
\thinspace $n-$ary code $V,$ is

$M=\left( 
\begin{tabular}{|l|l|l|l|l|l|l|l|l|}
\hline
$0$ & $0$ & $0$ & $0$ & $0$ & $0$ & $0$ & $0$ & $0$ \\ \hline
$1$ & $0$ & $0$ & $0$ & $0$ & $0$ & $0$ & $0$ & $0$ \\ \hline
$2$ & $1$ & $0$ & $0$ & $0$ & $0$ & $0$ & $0$ & $0$ \\ \hline
$3$ & $1$ & $1$ & $0$ & $0$ & $0$ & $0$ & $0$ & $0$ \\ \hline
$4$ & $1$ & $1$ & $1$ & $0$ & $0$ & $0$ & $0$ & $0$ \\ \hline
$5$ & $1$ & $1$ & $1$ & $1$ & $0$ & $0$ & $0$ & $0$ \\ \hline
$6$ & $\mathbf{3}$ & $\mathbf{2}$ & $\mathbf{1}$ & $\mathbf{1}$ & $1$ & $0$
& $0$ & $0$ \\ \hline
$7$ & $\mathbf{4}$ & $\mathbf{2}$ & $\mathbf{2}$ & $\mathbf{1}$ & $1$ & $1$
& $0$ & $0$ \\ \hline
$8$ & $\mathbf{4}$ & $\mathbf{3}$ & $\mathbf{2}$ & $\mathbf{1}$ & $1$ & $1$
& $1$ & $0$ \\ \hline
\end{tabular}%
\right) $ \newline

and the corresponded $BCK$-algebra, $\left( X,\ast ,\theta \right) ,$ where

$X=\{a_{0}=\theta ,a_{1},a_{2},a_{3},a_{4},a_{5},a_{6},a_{7},a_{8}\},$

with the following multiplication table

\begin{tabular}{l|l|l|l|l|l|l|l|l|l|}
$\ast $ & $\theta $ & $a_{1}$ & $a_{2}$ & $a_{3}$ & $a_{4}$ & $a_{5}$ & $%
a_{6}$ & $a_{7}$ & $a_{8}$ \\ \hline
$\theta $ & $\theta $ & $\theta $ & $\theta $ & $\theta $ & $\theta $ & $%
\theta $ & $\theta $ & $\theta $ & $\theta $ \\ \hline
$a_{1}$ & $a_{1}$ & $\theta $ & $\theta $ & $\theta $ & $\theta $ & $\theta $
& $\theta $ & $\theta $ & $\theta $ \\ \hline
$a_{2}$ & $a_{2}$ & $a_{1}$ & $\theta $ & $\theta $ & $\theta $ & $\theta $
& $\theta $ & $\theta $ & $\theta $ \\ \hline
$a_{3}$ & $a_{3}$ & $a_{1}$ & $a_{1}$ & $\theta $ & $\theta $ & $\theta $ & $%
\theta $ & $\theta $ & $\theta $ \\ \hline
$a_{4}$ & $a_{4}$ & $a_{1}$ & $a_{1}$ & $a_{1}$ & $\theta $ & $\theta $ & $%
\theta $ & $\theta $ & $\theta $ \\ \hline
$a_{5}$ & $a_{5}$ & $a_{1}$ & $a_{1}$ & $a_{1}$ & $a_{1}$ & $\theta $ & $%
\theta $ & $\theta $ & $\theta $ \\ \hline
$a_{6}$ & $a_{6}$ & $\mathbf{a}_{3}$ & $\mathbf{a}_{2}$ & $\mathbf{a}_{1}$ & 
$\mathbf{a}_{1}$ & $a_{1}$ & $\theta $ & $\theta $ & $\theta $ \\ \hline
$a_{7}$ & $a_{7}$ & $\mathbf{a}_{4}$ & $\mathbf{a}_{2}$ & $\mathbf{a}_{2}$ & 
$\mathbf{a}_{1}$ & $a_{1}$ & $a_{1}$ & $\theta $ & $\theta $ \\ \hline
$a_{8}$ & $a_{8}$ & $\mathbf{a}_{4}$ & $\mathbf{a}_{3}$ & $\mathbf{a}_{2}$ & 
$\mathbf{a}_{1}$ & $a_{1}$ & $a_{1}$ & $a_{1}$ & $\theta $ \\ \hline
\end{tabular}%
\newline

If we consider $A=\{1,2,3,4\}.$ The map $f:A\rightarrow X,$ $f\left(
1\right) =a_{1},f\left( 2\right) =a_{2},f\left( 3\right) =a_{3},f\left(
4\right) =a_{4}$ gives us the following block code\newline
$V^{\prime }=\{0000,1000,1100,1110,1111,\mathbf{3211,4221,4321}\},$ which
contains $V~$\ as a subset.

We remark that this algebra is not commutative since $a_{7}\ast \left(
a_{7}\ast a_{6}\right) =a_{7}\ast a_{1}=a_{4}$ and $a_{6}\ast (a_{6}\ast
a_{7})=a_{6}\ast \theta =a_{6}.$ This algebra is not implicative since $%
a_{6}\ast \left( a_{7}\ast a_{6}\right) =a_{6}\ast a_{1}=a_{3}\neq a_{6}.$
This algebra is not positive implicative since $\left( x\ast y\right) \ast
z\neq \left( x\ast z\right) \ast (y\ast z).$ Indeed, $\left( a_{7}\ast
a_{6}\right) \ast a_{3}=a_{1}\ast a_{3}=\theta \neq \left( a_{7}\ast
a_{3}\right) \ast \left( a_{6}\ast a_{3}\right) =a_{2}\ast
a_{1}=a_{1}.\medskip $

\textbf{Example 3.2.} Let $A_{4}=\{0,1,2,3\},$ $n=4,$ $q=5,$ $m=3,$ $r=9,$ $%
V=\{w_{1},w_{2},w_{3}\},$ with $w_{1}=21111,w_{2}=32111,w_{3}=33111.$ We
obtain the matrix $M$ associated to the \thinspace $n-$ary code $V,$

$M=\left( 
\begin{tabular}{|l|l|l|l|l|l|l|l|l|}
\hline
$0$ & $0$ & $0$ & $0$ & $0$ & $0$ & $0$ & $0$ & $0$ \\ \hline
$1$ & $0$ & $0$ & $0$ & $0$ & $0$ & $0$ & $0$ & $0$ \\ \hline
$2$ & $1$ & $0$ & $0$ & $0$ & $0$ & $0$ & $0$ & $0$ \\ \hline
$3$ & $1$ & $1$ & $0$ & $0$ & $0$ & $0$ & $0$ & $0$ \\ \hline
$4$ & $1$ & $1$ & $1$ & $0$ & $0$ & $0$ & $0$ & $0$ \\ \hline
$5$ & $1$ & $1$ & $1$ & $1$ & $0$ & $0$ & $0$ & $0$ \\ \hline
$6$ & $\mathbf{2}$ & $\mathbf{1}$ & $\mathbf{1}$ & $\mathbf{1}$ & $\mathbf{1}
$ & $0$ & $0$ & $0$ \\ \hline
$7$ & $\mathbf{3}$ & $\mathbf{2}$ & $\mathbf{1}$ & $\mathbf{1}$ & $\mathbf{1}
$ & $1$ & $0$ & $0$ \\ \hline
$8$ & $\mathbf{3}$ & $\mathbf{3}$ & $\mathbf{1}$ & $\mathbf{1}$ & $\mathbf{1}
$ & $1$ & $1$ & $0$ \\ \hline
\end{tabular}%
\right) $ \newline

and the corresponded $BCK$-algebra, $\left( X,\ast ,\theta \right) ,$ where

$X=\{a_{0}=\theta ,a_{1},a_{2},a_{3},a_{4},a_{5},a_{6},a_{7},a_{8}\},$

with the following multiplication table

\begin{tabular}{l|l|l|l|l|l|l|l|l|l|}
$\ast $ & $\theta $ & $a_{1}$ & $a_{2}$ & $a_{3}$ & $a_{4}$ & $a_{5}$ & $%
a_{6}$ & $a_{7}$ & $a_{8}$ \\ \hline
$\theta $ & $\theta $ & $\theta $ & $\theta $ & $\theta $ & $\theta $ & $%
\theta $ & $\theta $ & $\theta $ & $\theta $ \\ \hline
$a_{1}$ & $a_{1}$ & $\theta $ & $\theta $ & $\theta $ & $\theta $ & $\theta $
& $\theta $ & $\theta $ & $\theta $ \\ \hline
$a_{2}$ & $a_{2}$ & $a_{1}$ & $\theta $ & $\theta $ & $\theta $ & $\theta $
& $\theta $ & $\theta $ & $\theta $ \\ \hline
$a_{3}$ & $a_{3}$ & $a_{1}$ & $a_{1}$ & $\theta $ & $\theta $ & $\theta $ & $%
\theta $ & $\theta $ & $\theta $ \\ \hline
$a_{4}$ & $a_{4}$ & $a_{1}$ & $a_{1}$ & $a_{1}$ & $\theta $ & $\theta $ & $%
\theta $ & $\theta $ & $\theta $ \\ \hline
$a_{5}$ & $a_{5}$ & $a_{1}$ & $a_{1}$ & $a_{1}$ & $a_{1}$ & $\theta $ & $%
\theta $ & $\theta $ & $\theta $ \\ \hline
$a_{6}$ & $a_{6}$ & $\mathbf{a}_{2}$ & $\mathbf{a}_{1}$ & $\mathbf{a}_{1}$ & 
$\mathbf{a}_{1}$ & $\mathbf{a}_{1}$ & $\theta $ & $\theta $ & $\theta $ \\ 
\hline
$a_{7}$ & $a_{7}$ & $\mathbf{a}_{3}$ & $\mathbf{a}_{2}$ & $\mathbf{a}_{2}$ & 
$\mathbf{a}_{1}$ & $\mathbf{a}_{1}$ & $a_{1}$ & $\theta $ & $\theta $ \\ 
\hline
$a_{8}$ & $a_{8}$ & $\mathbf{a}_{3}$ & $\mathbf{a}_{3}$ & $\mathbf{a}_{1}$ & 
$\mathbf{a}_{1}$ & $\mathbf{a}_{1}$ & $a_{1}$ & $a_{1}$ & $\theta $ \\ \hline
\end{tabular}%
\newline

If we consider $A=\{1,2,3,4,5\}.$ The map $f:A\rightarrow X,$ $f\left(
1\right) =a_{1},f\left( 2\right) =a_{2},f\left( 3\right) =a_{3},f\left(
a_{4}\right) =4,f\left( a_{5}\right) =5,$ gives us the following block code $%
V_{X}=\{00000,10000,11000,11100,11110,\mathbf{21111,32211,33111}\},$ which
contains $V~$\ as a subset.\medskip

\textbf{Example 3.3.} We consider $A_{4}=\{0,1,2,3\},$ $n=4,$ $q=5,$ $m=5,$ $%
r=11,$ $V=\{w_{1},w_{2},w_{3},w_{4},w_{5}\},$ with $%
w_{1}=11111,w_{2}=21111,w_{3}=31111,w_{4}=32111,w_{5}=33111.$ We obtain the
matrix $M$ associated to the \thinspace $n-$ary code $V,$

$M=\left( 
\begin{tabular}{|l|l|l|l|l|l|l|l|l|l|l|}
\hline
$0$ & $0$ & $0$ & $0$ & $0$ & $0$ & $0$ & $0$ & $0$ & $0$ & $0$ \\ \hline
$1$ & $0$ & $0$ & $0$ & $0$ & $0$ & $0$ & $0$ & $0$ & $0$ & $0$ \\ \hline
$2$ & $1$ & $0$ & $0$ & $0$ & $0$ & $0$ & $0$ & $0$ & $0$ & $0$ \\ \hline
$3$ & $1$ & $1$ & $0$ & $0$ & $0$ & $0$ & $0$ & $0$ & $0$ & $0$ \\ \hline
$4$ & $1$ & $1$ & $1$ & $0$ & $0$ & $0$ & $0$ & $0$ & $0$ & $0$ \\ \hline
$5$ & $1$ & $1$ & $1$ & $1$ & $0$ & $0$ & $0$ & $0$ & $0$ & $0$ \\ \hline
$6$ & $\mathbf{1}$ & $\mathbf{1}$ & $\mathbf{1}$ & $\mathbf{1}$ & $\mathbf{1}
$ & $0$ & $0$ & $0$ & $0$ & $0$ \\ \hline
$7$ & $\mathbf{2}$ & $\mathbf{1}$ & $\mathbf{1}$ & $\mathbf{1}$ & $\mathbf{1}
$ & $1$ & $0$ & $0$ & $0$ & $0$ \\ \hline
$8$ & $\mathbf{3}$ & $\mathbf{1}$ & $\mathbf{1}$ & $\mathbf{1}$ & $\mathbf{1}
$ & $1$ & $1$ & $0$ & $0$ & $0$ \\ \hline
$9$ & $\mathbf{3}$ & $\mathbf{2}$ & $\mathbf{1}$ & $\mathbf{1}$ & $\mathbf{1}
$ & $1$ & $1$ & $1$ & $0$ & $0$ \\ \hline
$10$ & $\mathbf{3}$ & $\mathbf{3}$ & $\mathbf{1}$ & $\mathbf{1}$ & $\mathbf{1%
}$ & $1$ & $1$ & $1$ & $1$ & $0$ \\ \hline
\end{tabular}%
\right) $ \newline

and the corresponded $BCK$-algebra, $\left( X,\ast ,\theta \right) ,$ where

$X=\{a_{0}=\theta
,a_{1},a_{2},a_{3},a_{4},a_{5},a_{6},a_{7},a_{8},a_{9},a_{10}\},$

with the following multiplication table

\begin{tabular}{l|l|l|l|l|l|l|l|l|l|l|l|}
$\ast $ & $\theta $ & $a_{1}$ & $a_{2}$ & $a_{3}$ & $a_{4}$ & $a_{5}$ & $%
a_{6}$ & $a_{7}$ & $a_{8}$ & $a_{9}$ & $a_{10}$ \\ \hline
$\theta $ & $\theta $ & $\theta $ & $\theta $ & $\theta $ & $\theta $ & $%
\theta $ & $\theta $ & $\theta $ & $\theta $ & $\theta $ & $\theta $ \\ 
\hline
$a_{1}$ & $a_{1}$ & $\theta $ & $\theta $ & $\theta $ & $\theta $ & $\theta $
& $\theta $ & $\theta $ & $\theta $ & $\theta $ & $\theta $ \\ \hline
$a_{2}$ & $a_{2}$ & $a_{1}$ & $\theta $ & $\theta $ & $\theta $ & $\theta $
& $\theta $ & $\theta $ & $\theta $ & $\theta $ & $\theta $ \\ \hline
$a_{3}$ & $a_{3}$ & $a_{1}$ & $a_{1}$ & $\theta $ & $\theta $ & $\theta $ & $%
\theta $ & $\theta $ & $\theta $ & $\theta $ & $\theta $ \\ \hline
$a_{4}$ & $a_{4}$ & $a_{1}$ & $a_{1}$ & $a_{1}$ & $\theta $ & $\theta $ & $%
\theta $ & $\theta $ & $\theta $ & $\theta $ & $\theta $ \\ \hline
$a_{5}$ & $a_{5}$ & $a_{1}$ & $a_{1}$ & $a_{1}$ & $a_{1}$ & $\theta $ & $%
\theta $ & $\theta $ & $\theta $ & $\theta $ & $\theta $ \\ \hline
$a_{6}$ & $a_{6}$ & $\mathbf{a}_{1}$ & $\mathbf{a}_{1}$ & $\mathbf{a}_{1}$ & 
$\mathbf{a}_{1}$ & $\mathbf{a}_{1}$ & $\theta $ & $\theta $ & $\theta $ & $%
\theta $ & $\theta $ \\ \hline
$a_{7}$ & $a_{7}$ & $\mathbf{a}_{2}$ & $\mathbf{a}_{1}$ & $\mathbf{a}_{1}$ & 
$\mathbf{a}_{1}$ & $\mathbf{a}_{1}$ & $a_{1}$ & $\theta $ & $\theta $ & $%
\theta $ & $\theta $ \\ \hline
$a_{8}$ & $a_{8}$ & $\mathbf{a}_{3}$ & $\mathbf{a}_{1}$ & $\mathbf{a}_{1}$ & 
$\mathbf{a}_{1}$ & $\mathbf{a}_{1}$ & $a_{1}$ & $a_{1}$ & $\theta $ & $%
\theta $ & $\theta $ \\ \hline
$a_{9}$ & $a_{9}$ & $\mathbf{a}_{3}$ & $\mathbf{a}_{2}$ & $\mathbf{a}_{1}$ & 
$\mathbf{a}_{1}$ & $\mathbf{a}_{1}$ & $a_{1}$ & $a_{1}$ & $a_{1}$ & $\theta $
& $\theta $ \\ \hline
$a_{10}$ & $a_{10}$ & $\mathbf{a}_{3}$ & $\mathbf{a}_{3}$ & $\mathbf{a}_{1}$
& $\mathbf{a}_{1}$ & $\mathbf{a}_{1}$ & $a_{1}$ & $a_{1}$ & $a_{1}$ & $a_{1}$
& $\theta $ \\ \hline
\end{tabular}%
\newline

If we consider $A=\{1,2,3,4,5\}.$ The map $f:A\rightarrow X,$ $f\left(
1\right) =a_{1},f\left( 2\right) =a_{2},f\left( 3\right) =a_{3},f\left(
a_{4}\right) =4,f\left( a_{5}\right) =5,$ gives us the following block code $%
V^{\prime }=\{00000,10000,11000,11100,11110,\mathbf{%
11111,21111,31111,32111,33111}\},$ which contains $V~$\ as a subset.\medskip

\bigskip

\textbf{Conclusions.} In this paper, we \ proved that to each $n-$ary block
code $V$ we can associate a $BCK$-algebra $X$ such that the $n-$ary
block-code generated by $X,V_{X},$ contains the code $V$ as a subset. This
algebra is unique up to an isomorphism and $X$ is not commutative, not
implicative and not positive implicative $BCK$-algebra.

As a further research will be very interesting to study properties of the
above constructed codes and how these codes in connections with their
associated $BCK$-algebras.

\begin{equation*}
\end{equation*}

\textbf{References}%
\begin{equation*}
\end{equation*}

[B,F; 15] Borumand Saeid, A., Fatemidokht, H., Flaut, C., Kuchaki
Rafsanjani, M., \textit{On Codes based on BCK-algebras}, Journal of
Intelligent and Fuzzy Systems, 29 (2015) 2133--2137,

[Fl; 15] C. Flaut, \textit{BCK-algebras arising from block codes}, Journal
of Intelligent and Fuzzy Systems, 28(4) (2015), 1829--1833.

[Im, Is; 66] Y. Imai, K. Iseki, \textit{On axiom systems of propositional
calculi}, Proc. Japan Academic, 42 (1966), 19-22.

[Is, Ta; 78] K. Iseki and S. Tanaka,\textit{\ An introduction to the theory
of BCK-algebras}, Math Jpn, 23 (1978), 1--26.

[Ju,So; 11] Y. B. Jun and S. Z. Song, \textit{Codes based on BCK-algebras,}
Inform Sciences, 181 (2011), 5102--5109.

[Me, Ju; 94] J. Meng, Y. B. Jun, \textit{BCK-algebras,} Kyung Moon, \ Seoul,
Korea, 1994.

\begin{equation*}
\end{equation*}
\begin{equation*}
\end{equation*}
\begin{equation*}
\end{equation*}

Arsham Borumand Saeid

{\small Department of Pure Mathematics,}

{\small Faculty of Mathematics and Computer,}

{\small Shahid Bahonar University of Kerman, Kerman, Iran }

{\small e-mail: arsham@uk.ac.ir}

\bigskip

Cristina Flaut

{\small Faculty of Mathematics and Computer Science, Ovidius University,}

{\small Bd. Mamaia 124, 900527, CONSTANTA, ROMANIA}

{\small http://cristinaflaut.wikispaces.com/;
http://www.univ-ovidius.ro/math/}

{\small e-mail: cflaut@univ-ovidius.ro; cristina\_flaut@yahoo.com}

\bigskip

Sarka Ho\v{s}kov\'{a}-Mayerov\'{a}

{\small University of Defence, Brno}

{\small sarka.mayerova@unob.cz}

\bigskip

Roxana-Lavinia Cristea

{\small Project manager \& Mobile applications developer at Appscend
(www.appscend.com)}

{\small E-mail1: roxana@appscend.com}

{\small E-mail2: rox.lavinia@gmail.com}

\bigskip Morteza Afshar

{\small Department of Pure Mathematics,}

{\small Faculty of Mathematics and Computer,}

{\small Shahid Bahonar University of Kerman, Kerman, Iran }

\bigskip

Marjan Kuchaki Rafsanjani {\small Department of Computer Science,}

{\small Faculty of Mathematics and Computer,}

{\small Shahid Bahonar University of Kerman, Kerman, Iran }

{\small e-mail: kuchaki@uk.ac.ir}

\end{document}